\documentclass[fleqn,10pt,twocolumn]{wlscirep}
\usepackage{graphicx}    
\usepackage{amsmath}    
\usepackage{amssymb}    
\usepackage{comment}
\usepackage[export]{adjustbox}
\usepackage{verbatim}
\usepackage{mathrsfs}
\usepackage{authblk}
\usepackage{subfigure}
\usepackage[T1]{fontenc}
\usepackage{subfigure}
\usepackage{ae,aecompl}
\usepackage{xcolor}
\usepackage{marginnote}  
\usepackage{soul}        
\usepackage{txfonts}

\title{Magnetic Field Morphology in Interstellar Clouds with the Velocity Gradient Technique}
\author[1,2]{Yue Hu}
\author[2]{Ka Ho Yuen}
\author[2]{Victor Lazarian}
\author[3]{Ka Wai Ho}
\author[2]{Robert A. Benjamin}
\author[4,5,6]{Alex S. Hill}
\author[7]{Felix J. Lockman}
\author[8]{Paul F. Goldsmith}
\author[2]{A. Lazarian}	
\affil[1]{Department of Physics, University of Wisconsin-Madison,USA}
\affil[2]{Department of Astronomy, University of Wisconsin-Madison,USA}
\affil[3]{Department of Physics, The Chinese University of Hong Kong, Hong Kong}
\affil[4]{Department of Physics and Astronomy, University of British Columbia, Vancouver, BC Canada}
\affil[5]{Space Science Institute, Boulder, CO USA}
\affil[6]{Dominion Radio Astrophysical Observatory, National Research Council Canada, Penticton, BC Canada}
\affil[7]{Green Bank Observatory, P.O. Box 2, Route 28/92, Green Bank, WV 24944, USA}
\affil[8]{Jet Propulsion Laboratory, California Institute of Technology, USA}

\begin{abstract}
	Magnetic fields, while ubiquitous in many astrophysical  environments, are challenging to measure observationally. Based on the properties of anisotropy of eddies in magnetized turbulence, the Velocity Gradient Technique is a method synergistic to dust polarimetry that is capable of tracing plane-of-the-sky magnetic field, measuring the magnetization of interstellar media and estimating the fraction of gravitational collapsing gas in molecular clouds using spectral line observations. In this paper, we apply this technique to five low-mass star-forming molecular clouds in the Gould Belt and compare the results to the magnetic-field orientation obtained from  polarized dust emission. We find the estimates of magnetic field orientations and magnetization for both methods are statistically similar.  We estimate the fraction of collapsing gas in the selected clouds. By means of the Velocity Gradient Technique, we also present the plane-of-the-sky magnetic field orientation and magnetization of the Smith cloud, for which dust polarimetry data are unavailable.
\end{abstract}
\begin{document}

\flushbottom
\maketitle
%
%
\thispagestyle{empty}


Magnetic fields play a crucial role in a variety of important astrophysical processes, from regulating molecular cloud structure formation and evolution\cite{2016AA...586A.138P,1981MNRAS.194..809L,2003ApJ...585..850M} to constraining star formation in filaments\cite{2015MNRAS.452.2410S,1999ApJ...527L...5B,2011AA...533A..94H}. 
The importance of the magnetic field can be characterized by the ratio of the kinetic energy $\sim 0.5\rho v_L^2$ to the magnetic energy $\sim B^2/8\pi\rho$ in the cloud, where $v_L$ is the turbulent velocity at scale $L$, $\rho$ is the volume density, and $B$ is the magnetic field strength.  This quantity is the inverse square of the Alfven Mach number (M$_A^{-1})^2=B^2/(4\pi v_L^2)$, a variable used in both theory of magnetic turbulence and cosmic ray propagation \cite{2013SSRv..178..163B}. 

Due to advances in dust grain alignment theory\cite{2015ARAA..53..501A}, the properties of magnetic fields have become more accessible using dust polarization from background starlight or polarized thermal dust emission. For example, the recent Planck survey of polarized dust emission provided us with a comprehensive picture of magnetic field orientations across the full sky\cite{aghanim2018planck}.  Similarly, dust polarimetric surveys have significantly advanced our knowledge of the magnetic field orientations in molecular clouds\cite{nature2015}. There are, however, challenges when studying the magnetic field through dust polarimetry. For one, dust polarimetry becomes ineffective in the case in which the grains are not aligned. Modern grain alignment theory\cite{2003JQSRT..79..881L} suggests that grain alignment is  driven mainly by radiative torques, but the grains become misaligned in a number of circumstances. For instance, in the absence of sufficiently intense radiation, the orientation of dust grains is random \cite{2015ARAA..53..501A}. While in the vicinity of radiation sources dust grains can be aligned with respect to the incident radiation rather than to the ambient magnetic field \cite{2007ApJ...669L..77L,2018ApJ...852..129H}. In addition, it is impossible to separate the contributions of overlapping molecular clouds because dust polarimetry measurements from millimeter, submillimeter, or far-infrared emission sample all the dust along the line of sight. The failure of radiative-torque-driven dust grain alignment mechanisms in high-extinction environments affects further predictions of magnetic field properties based on dust polarimetry measurements, e.g. using the Davis-Chandrasekhar-Fermi (DCF) technique\cite{1951PhRv...81..890D,1953ApJ...118..113C}.

Aside from dust polarization, there are alternative ways of probing the magnetic field structure. Zeeman measurements allow observers to estimate the signed magnetic field strength along the line of sight and have contributed significantly to our understanding of star formation \cite{2010ApJ...725..466C}. However, these measurements require extremely high sensitivity and long integration times. In addition, usually only upper limits of the magnetic field strength are obtained by the Zeeman method. Another tool used to measure the magnetic field strength along the line of sight is Faraday Rotation towards polarized radio point sources \cite{2009ApJ...702.1230T}. Faraday rotation measures the electron density-weighted magnetic field strength along the line of sight and therefore generally does not probe the magnetic field in primarily neutral regions such as molecular clouds. Therefore, there is a demand for alternative methods for probing magnetic fields.

The Velocity Gradient Technique (VGT)\cite{2017ApJ...835...41G,2017ApJ...837L..24Y,2018ApJ...853...96L,2018ApJ...865...54Y} is a new method capable of tracing the magnetic field orientation in interstellar turbulent media. The technique makes use of the fact that magnetohydrodynamic (MHD) turbulence is anisotropic\cite{1995ApJ...438..763G}. It is important that fast turbulent reconnection, the process by which magnetic fields in a conducting fluid change their topology driven by turbulence and independent of fluid resistivity, preferentially induces fluid motions perpendicular to the local magnetic field direction\cite{1999ApJ...517..700L}. As a result, gradients of velocities become perpendicular to the local direction of the magnetic field. This phenomenon has been numerically confirmed\cite{2000ApJ...539..273C,2001ApJ...554.1175M,2010ApJ...722L.110B,beresnyak2015mhd} and is the cornerstone of the modern theory of MHD turbulence\cite{2013SSRv..178..163B}.

\begin{table*}[htb]
\centering
\begin{tabular}{| c | c | c | c | c | c | c |}
\hline
Cloud Region & Taurus & Perseus A& L 1551& Serpens & NGC 1333  & Smith Cloud\\\hline
Emission lines & $^{13}$CO: J=1-0 & $^{13}$CO: J=1-0 & $^{13}$CO: J=1-0& $^{13}$CO: J=2-1 & $^{13}$CO: J=2-1  & H I 21cm\\\hline
AM & 0.74$\pm$0.03 & 0.52$\pm$0.05& 0.70$\pm$0.03 & -0.81$\pm$0.03 & -0.71$\pm$0.03  & ...\\\hline
M$_A$ & 1.19$\pm$0.02 & 1.22$\pm$0.05 & 0.73$\pm$0.13& 0.98$\pm$0.08 & 0.82$\pm$0.24  & 0.68$\pm$0.12\\\hline
M$_A^P$ & 1.13$\pm$0.02 & 1.20$\pm$0.02& 0.67$\pm$0.05 & 0.78$\pm$0.03 & 0.95$\pm$0.03 & ...\\\hline
$\mu$ & 86.12$^\circ\pm$1.21$^\circ$ & 88.72$^\circ\pm$1.08$^\circ$ & 85.10$^\circ\pm$1.95$^\circ$& 10.06$^\circ\pm$1.42$^\circ$ & 8.08$^\circ\pm$1.41$^\circ$  & ... \\\hline
\end{tabular}
\caption{\label{tab:data} \textbf{Information about the regions and data used in this work. } All quantities are averages over the regions. 
	Terminology: AM: Alignment Measure, where AM = 1 indicates perfect alignment between VGT and Planck polarization vectors;   M$_A$: Alfv\'{e}nic Mach number derived from VGT; M$_A^P$: Alfv\'{e}nic Mach number derived from Planck polarization. $\mu$ is the expectation of the relative angle between the un-rotated gradients and the magnetic field derived from Planck polarization. The uncertainty is given by the standard error of the mean, i.e. the standard deviation divided by the square root of the sample size (see Supplementary Information for details).}
\end{table*} 

The VGT has been numerically tested for a wide range of column densities from diffuse transparent gas \cite{2017ApJ...837L..24Y,2019ApJ...886...17H,2020ApJ...888...96H} to molecular self-absorbing dense gas \cite{2019ApJ...873...16H}. It was shown to be able to provide both the  orientations of the magnetic field as well as a measure of media magnetization \cite{2018ApJ...865...46L,2019arXiv190404391H}. The technique has been used to study magnetic field in diffuse H I \cite{2017ApJ...837L..24Y,2018ApJ...853...96L,2018ApJ...865...46L,2018MNRAS.480.1333H} and was shown to be complementary to other methods of tracing magnetic field\cite{2018ApJ...865...54Y}. This paper is the application of the VGT to molecular clouds which are known to be turbulent \cite{1981MNRAS.194..809L,2001ApJ...546..980O} and magnetized \cite{2010ApJ...725..466C}. While some of the structure in the spectroscopic data may not be due to MHD turbulence, recent studies of VGT show the way of identifying these situations and using them to study other important interstellar physics, e.g. the gravitational collapse of a cloud of interstellar matter \cite{2018ApJ...865...46L}.

In this paper, we apply the VGT to five low mass molecular clouds. The case of massive star formation clouds in which the effects of gravitational collapse are more significant will be investigated elsewhere. This work studies magnetic fields on the scales at which ions and neutrals are well coupled\cite{2017NJPh...19f5005X} and therefore our expectations based on MHD turbulence theory are applicable. We compare our results with the 353 $GHz$  polarization data from the 3rd Public Data Release (DR3) from the Planck Collaboration in 2018 \cite{aghanim2018planck}. In addition, to illustrate the abilities of the technique, we present our prediction of magnetic field orientations and magnetization for the Smith Cloud\cite{1963BAN....17..203S}, a magnetized high-velocity cloud of atomic hydrogen falling into the Milky Way, for which no optical or infrared polarimetric data are available \cite{2013ApJ...777...55H,2019ApJ...871..215B}.


\section*{Results}
\subsection*{Morphology of Magnetic Field Traced by The Velocity Gradient Technique}

\begin{figure*}[!h]
\centering
\includegraphics[width=1.00\linewidth,height=0.75\linewidth]{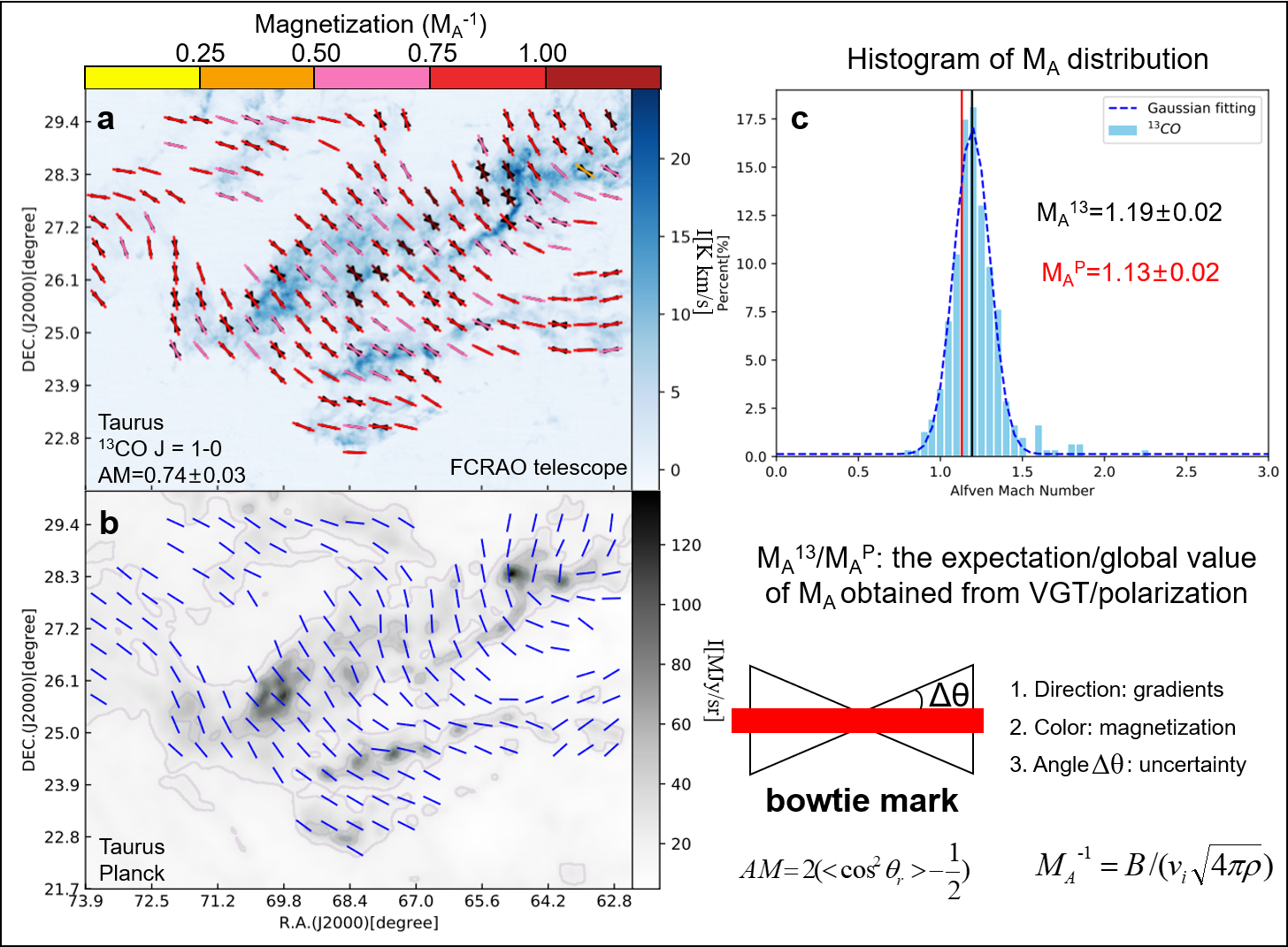}
\caption{The magnetic field morphology of Taurus obtained with the VGT using $^{13}$CO and the Planck polarimetry. \textbf{Panel a}, the magnetic field morphology of Taurus obtained with the VGT using $^{13}$CO. \textbf{Panel b}, the blue line segments indicate the average orientation of the magnetic field the magnetic field morphology of Taurus obtained from the Planck polarimetry. \textbf{Panel c}, the histogram of the M$_A$ distribution obtained with VGT on Taurus. As the VGT provides both the magnetic field orientation and its magnetization, we introduce a new symbol (right bottom), the bowtie, which reflects the magnetic field orientation, and magnetization, i.e. M$_A^{-1}$, (by the coloured segment, different colors corresponding to different magnetization) and the dispersion of orientations (by the angle of the bowtie) that follows from the VGT. M$_{A}^{P}$ (red) represents the mean M$_A$ value obtained from the dispersion of polarization orientations, while M$_{A}^{13}$ is the expectation value of M$_A$ obtained from VGT.}
\label{fig:survey}
\end{figure*}

\begin{figure*}[!h]
\centering
\includegraphics[width=1.00\linewidth,height=0.6\linewidth]{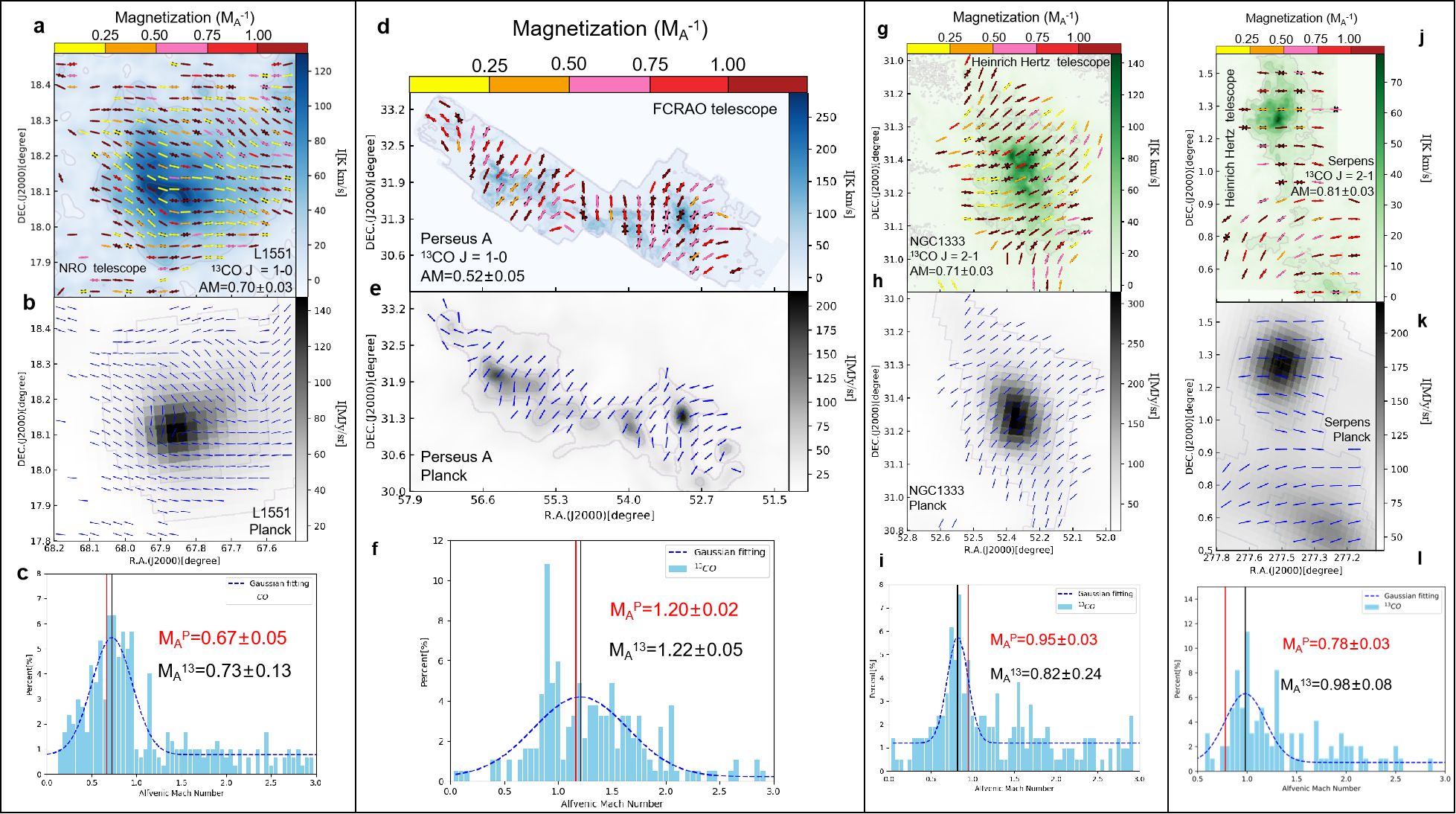}
\caption{The magnetic field morphology of molecular clouds L 1551 (the 1$^{st}$ column), Perseus A (the 2$^{nd}$ column), NGC 1333 (the 3$^{rd}$ column), and Serpens (the 4$^{th}$ column) obtained with the VGT using $^{13}$CO and the Planck polarimetry. \textbf{Panel a, d, g, and j}, the magnetic field morphology of L 1551 (\textbf{a}), Perseus A (\textbf{d}), NGC 1333 (\textbf{g}), and Serpens (\textbf{j}) respectively obtained with the VGT using $^{13}$CO (the gradients on NGC 1333 and Serpens are re-rotated). \textbf{Panel b, e, h, k}, the blue line segments indicate the average orientation of the magnetic field the magnetic field morphology of L 1551 (\textbf{b}), Perseus A (\textbf{e}), NGC 1333 (\textbf{h}), and Serpens (\textbf{k}) respectively obtained from the Planck polarimetry. \textbf{Panel c, f, i, and l} are the histograms of the M$_A$ distribution obtained with VGT on  L 1551 (\textbf{c}), Perseus A (\textbf{f}), NGC 1333 (\textbf{i}), and Serpens (\textbf{l}) respectively.  We use the same bowtie mark to indicate the orientation of the magnetic field and the magnetization as in Fig.~\ref{fig:survey}. M$_{A}^{P}$ represents the mean M$_A$ value obtained from the dispersion of polarization orientations, while M$_{A}^{13}$ is the expectation value of M$_A$ obtained from VGT.}
\label{fig:survey2}
\end{figure*}
\begin{figure*}[!h]
\centering
\includegraphics[width=0.83\linewidth,height=0.65\linewidth]{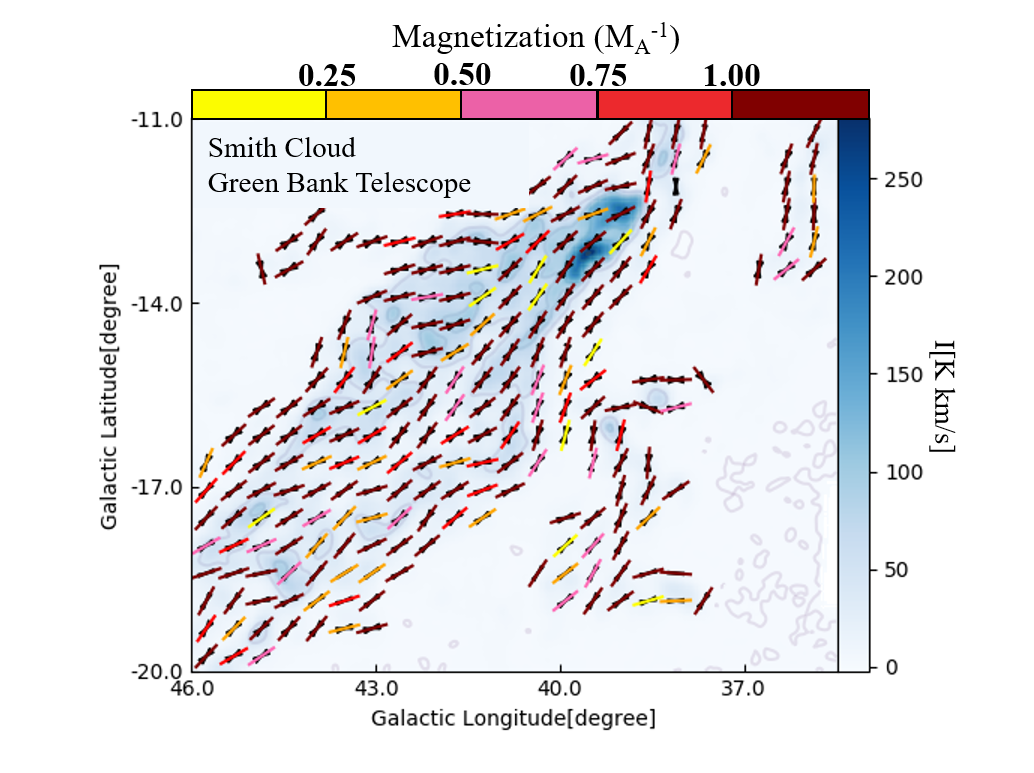}
\caption{ The magnetic field morphology and magnetization of the Smith Cloud  superimposed on a map of its total H I intensity. We use the same bowtie mark to indicate the orientation of the magnetic field and magnetization as shown in Fig.~\ref{fig:survey}. The color scale shows the integral of the H I line brightness temperature over velocity.}
\label{fig:susan}
\end{figure*}


The molecular clouds to which we apply VGT are: Taurus\cite{2008ApJ...680..428G}, Perseus A\cite{2006AJ....131.2921R}, L 1551\cite{2016ApJ...826..193L}, NGC1333\cite{2014ApJS..214....7B} and Serpens\cite{2013ApJS..209...39B}.  We use the spectroscopic maps of the molecular tracer $^{13}$CO from these molecular clouds to explore the plane-of-the-sky morphology of the magnetic field. The $^{13}$CO data on Taurus and Perseus A were obtained using the 13.7m Five College Radio Astronomy Observatory telescope, for L1551 the Nobeyama Radio Observatory 45 m telescope,  and those on NGC133 and Serpens were obtained with the Arizona Radio Observatory Heinrich Hertz Submillimeter Telescope. We compare our results to the Planck 353~$GHz$ polarization maps \cite{aghanim2018planck}, which provide the best representation of plane-of-the-sky magnetic field orientation of the aforementioned clouds that is available to us. Note that we do not smooth the data from Planck. Instead, we re-sample the Planck data so that the effective resolution matches that of the VGT-predicted magnetic field orientation (See Supplementary Table 1 for the effective resolution). The correspondence of the plane-of-the-sky magnetic field orientations obtained by the VGT and the polarization measurements is quantified using the  \textbf{Alignment Measure (AM)}: $AM=2(\langle cos^{2} \theta_{r}\rangle-\frac{1}{2})$, where $\theta_{r}$ is the relative angle between the gradients (rotated by 90$^o$) and the orientations of the plane-of-the-sky magnetic field derived from polarization. If the two measures provide identical results, $AM = 1$.

We adopt the recipe introduced by Lazarian \& Yuen (2018) \cite{2018ApJ...853...96L} to trace the plane-of-the-sky orientation of the magnetic field using the {\it {\bf G}radients of thin {\bf V}elocity {\bf Ch}annels} (VChGs, \textbf{Ch(x,y)}). The selection of thin channel maps (see Methods) increases the weight of velocity contribution to the measured statistics \cite{2000ApJ...537..720L}.  Due to properties of MHD turbulence\cite{2003MNRAS.345..325C}, velocity statistics trace the magnetic field orientation better than density statistics.

Fig.~\ref{fig:survey} shows the VChGs (rotated by 90$^\circ$) in Perseus A, while that in Fig.~\ref{fig:survey2} shows the results of VChGs for Taurus, L1551, NGC1333, and Serpens. For easy visual comparison we overlay the magnetic field orientations predicted by VChGs with those from dust polarization. From Fig.~\ref{fig:survey} and Fig.~\ref{fig:survey2}, we see that the magnetic field orientation predicted by VChGs and from dust polarimetry are in good agreement, with a mean statistical deviation of $\sim 4^\circ$. More detailed results of the statistical analysis are presented in Table~\ref{tab:data} (See the Supplementary Information for an analysis with the histogram of relative orientation and the alignment distribution map, i.e. Figs.~\ref{fig:HRO} and \ref{fig:AM}).

There are differences between the magnetic field orientations traced by the VGT and by the dust polarization observations, but that is expected. Spectroscopic data allow one to separate out different molecular clouds along the line of sight if they have different velocities, but this is not an option for polarized dust radiation. The VGT thus has an advantage for studying magnetic fields, especially for molecular clouds at low Galactic latitudes when the foreground and background polarization contributions are important. Other differences include the difference of the accumulation of the signal along the line of sight for gradients and polarization, which is especially important in the case of super-Alfvenic (M$_A>1$) turbulence \cite{2018ApJ...865...46L}. However, the good correlation between the VGT and the polarization measurement of magnetic field orientations shown in Table \ref{tab:data} demonstrates that all these factors are sub-dominant for the clouds at hand. However, Regions of gravitational collapse are expected to turn the direction of velocity gradients by 90$^\circ$ with respect to magnetic fields\cite{2017arXiv170303026Y}, which we will discuss in the following section.

Lazarian et al. (2018)\cite{2018ApJ...865...46L} demonstrated that the magnetization  parameterized by the inverse Alfven Mach number M$_A^{-1}$ can be estimated through the distribution of the velocity gradient orientations within the sub-blocks that are used within our technique (see Fig.~\ref{fig:dispersion}). In general, the distribution of the velocity gradient orientation is Gaussian \cite{2017ApJ...837L..24Y}. More importantly, the width of the distribution is shown to be correlated with the magnetization\cite{2018ApJ...865...46L}. In the case of magnetically dominated media (M$_A<1$), the distribution of velocity gradient orientations is narrower than in turbulence dominated media. We use the relations between the width of the gradient distribution and magnetization\cite{2018ApJ...865...46L} to evaluate the distribution of magnetization in the clouds that we study. Figs.~1 and 2 show the derived distribution of magnetization. For Taurus and Perseus A, we find that the magnetization in high-intensity regions is usually stronger than that in surrounding low-intensity regions. This corresponds to the increasing role of the magnetic field in regions of higher density.

We can use the distribution of polarization orientations over the entire region to obtain the mean magnetization (M$_A^P)^{-1}$. This measure involves the implicit use of Davis-Chandrasekhar-Fermi method. The correlation M$_A^P$ $\propto \tan\delta\theta_{pol}$ was numerically proven in Falceta-Goncalves et al. (2008)\cite{falceta2008studies}, where $\delta\theta_{pol}$ is the dispersion of the polarization angle. The mean magnetization can also be obtained by averaging of the local sub-block magnetization values obtained using the VGT. Table \ref{tab:data} illustrates the good correspondence between the two values. The advantage of the VGT compared to the traditional DCF technique is its ability to measure the not only mean magnetization but also a detailed distribution of the magnetization using a self-consistent algorithm \cite{2018ApJ...865...46L}, which is important for better understanding of star formation and other key astrophysical processes.

Neither polarization measurements nor the VGT trace magnetic fields in molecular clouds perfectly. For the Planck data, the measured signal includes not only the polarization from molecular clouds but also the contributions arising from aligned dust in the interstellar medium in front of and behind the cloud (See Supplementary information for the background removal). For VGT, the fast MHD modes present in MHD turbulence as well as the effects of gravitational collapse distort the directions of the gradients \cite{2018ApJ...853...96L}. In addition, VGT degrades the resolution of the spectroscopic maps due to the sub-block averaging method. As a trade off, however, one can get magnetization of the media for every sub-block. Importantly, the degradation of the resolution in the case of the VGT is compensated by the high resolution and good sampling of spectroscopic surveys as well as the abundance of available data. Therefore, the VGT is a valuable method for studying the magnetic field in a large variety of clouds for which no detailed polarization observations are available and in which individual clouds cannot be separated without velocity information.


\subsection*{Gravitational Collapse}
When turbulence is the dominant mechanism in the dynamics of molecular clouds, i.e. before self-gravity comes into play and where there is no distortion due to shocks and outflows, the velocity gradients are perpendicular to the local magnetic field. However, molecular clouds, in general, contain regions of gravitational collapse \cite{2015ApJ...804..141Z,2007prpl.conf...63B}. In the case with gravitational collapse, the infall motions parallel to the magnetic field will gradually dominate the velocity motions due to turbulence. When one measures the gradients of a highly self-gravitating molecular cloud, they follow the direction of the infall, i.e. the direction of gradients flips 90$^\circ$ becoming parallel to the magnetic field\cite{2017arXiv170303026Y,2020arXiv200206754H}. This happens because the acceleration of the fluid points toward the core of the collapsing region, thus the direction of the magnetic field and of fluid motions induced by gravitational infall become parallel. To account for this, compensatory re-rotation (i.e. rotating the gradients by additional 90$^\circ$) must be applied to the gradients\cite{2017arXiv170303026Y}. 

Fig.~\ref{fig:survey} shows the angle uncertainties of VChGs in Taurus, Fig.~\ref{fig:survey2} shows the uncertainties in Perseus A, L 1551, NGC 1333, and Serpens. We discuss in the Supplementary Information (see Fig.~\ref{fig:rerotation}) that our experimenting with re-rotating the gradient vectors and comparing the resulting gradient map with dust polarimetry indicates that there is no compensatory re-rotation required to obtain good alignment with magnetic fields derived from Planck on Taurus, L 1551, and Perseus A, but on NGC 1333 and Serpens.  We thus conclude that the gravitationally collapsing regions constitute only a small fraction of the volume in Taurus, L 1551, and Perseus A. This, however, does not prevent molecular gas in small regions (NGC 1333 and Serpens) from collapsing to form stars\cite{2011ApJ...734...63S}.

\subsection*{Prediction of Magnetic Field Morphology in the Smith Cloud}
We see from Table~\ref{tab:data} that the deviation between the average direction of the magnetic fields determined using the VGT and Planck polarimetry is within 5$^\circ$ for Taurus, Perseus~A and L~1551 , and within $10^\circ$ for Serpens, and NGC 1333. The demonstrated ability of VGT to trace the magnetic fields encourages us to apply it to interstellar clouds for which no dust polarimetry is available.

The Smith Cloud is a diffuse high-velocity H~{\sc I} cloud, and its radial velocity near $+100 \, \mathrm{km} \, \mathrm{s}^{-1}$ is inconsistent with Galactic rotation at its location \cite{1963BAN....17..203S,1998MNRAS.299..611B,2008ApJ...679L..21L}. Since the polarization orientation is dominated by the contribution of foreground Galactic media along the line-of-sight, dust polarization measurements are unlikely to reflect the magnetic field structure of the Smith Cloud\cite{2013ApJ...777...55H} itself. The advantage of the VGT is that it can provide information on the magnetic field structure of the cloud with limited contamination by the foreground Galactic emission. The VGT measures magnetic field in the plane of sky and is complementary to the measurements from previous studies \cite{2013ApJ...777...55H,2019ApJ...871..215B} that probe the line-of-sight component of the field in ionized gas by measuring the Faraday Rotation. 

Fig.~\ref{fig:susan} shows the predicted magnetic field orientations for the Smith Cloud using the VGT. The Smith Cloud is a diffuse cloud with no expected gravitational collapse. Therefore unlike molecular clouds, there should be no need for re-rotations of the velocity gradient orientations.  We find that the magnetization of the Smith Cloud is high: M$_A \approx 0.68\pm 0.12$ (see Fig.~\ref{fig:smith}). Since in highly magnetized regions the gradients show better alignment with the magnetic field\cite{2018ApJ...853...96L}, we expect the prediction of magnetic field morphology to be no less accurate than our results for the molecular clouds discussed previously.

Through parameters of the Smith cloud available from the literature\cite{2008ApJ...679L..21L,2013ApJ...777...55H,2008ApJ...672..298W}, we estimate the strength of the magnetic field $B > 3 \, \mu\mathrm{G}$ in Smith Cloud. The DCF method is used here, but instead of polarization we use the gradient orientation. The result is consistent with the estimate $B_{||} > 3 \, \mu\mathrm{G}$ by Hill et al. (2013)\cite{2013ApJ...777...55H} (see Supplementary Information for details).

\section*{Methods}
\label{method}
	\subsection*{The Velocity Gradient Technique}
The pioneering study by Goldreich \& Sridhar\cite{1995ApJ...438..763G}  opened a new era in the theory of MHD turbulence. Fast turbulent reconnection\cite{1999ApJ...517..700L}, the process by which magnetic fields in a conducting fluid change their topology  driven by turbulence and independent to fluid resistivity, is an important part of the modern understanding of the dynamics of the magnetized turbulent eddies that provide the theoretical foundations for the VGT. Lazarian \& Vishniac (1999) \cite{1999ApJ...517..700L} have predicted that magnetic field mixing motions that are perpendicular to the local magnetic field will be favored, since these kinds of motions will induce the least amount of magnetic field back-reaction. As a result, gradients of the fluid motions are expected to be perpendicular to the local magnetic field.

The VGT employs a statistical description of Position-Position-Velocity (PPV) cubes$^{\color{blue}{51}}$.  Lazarian \& Pogosyan (2000)\cite{2000ApJ...537..720L} explored the possibility of using the statistics of intensity fluctuations in PPV cubes to study velocity turbulence and the subsequent works used PPV cubes to detect the anisotropy of velocity distribution that is induced by the magnetic field $^{\color{blue}{52}}$. This velocity anisotropy is well-reflected in the preferred alignment of the intensity gradients measured in thin channel maps. In addition, the use of thin channels reduce the anonymous crowding effect due to both the overdensity and the opacity, which makes the gradients more aligned with the magnetic field\cite{2019ApJ...873...16H}. This is the approach that we use in our present study in order to trace the magnetic field.

The technique operates as follow: The velocity channel map \textbf{Ch(x,y)} is constructed by creating integrated maps over a narrow velocity range $\Delta v$ satisfying: $\Delta v \leq \sqrt{\delta v^{2}_{R}}$, where $\delta v^{2}_{R}$ is the velocity dispersion in a patch size of radius R. We choose the channel width of $\Delta v/\sqrt{\delta v^{2}_{R}}\sim 0.5$ so that the velocity contribution in the velocity channel map dominates over the density contribution\cite{2018ApJ...853...96L}.  We denote the selected velocity channels as  {\it thin}  channels, which can be calculated by integrating over velocity:
\begin{equation}
Ch(x,y)=\int_{\Delta v^{2} \leq \delta v^{2}_{R}} dv\ T_{R}(x,y,v)\cdot e^{-\frac{\mid v-v_{0}\mid ^{2}}{R^{2}}} ,
\end{equation}
where $T_{R}$ is the radiation temperature of the spectral line in units of kelvin (for H I data $T_{R}$ is 
proportional to the density), and $v$ is the line--of--sight velocity. From the velocity channel maps \textbf{Ch(x,y)}, the gradient orientation at pixel $(x_{i},y_{j})$ is defined as:
\begin{equation}
\bigtriangledown_{i,j}=tan^{-1}[\frac{Ch(x_{i},y_{j+1})-Ch(x_{i},y_{j})}{Ch(x_{i+1},y_{j})-Ch(x_{i},y_{j})}] .
\end{equation}
This creates the pixelized gradient orientation field for the spectroscopic data. When the velocity slice is thin, the channels  record the contribution of turbulent velocities\cite{2018ApJ...853...96L}. Thus the Velocity Channel Gradients (VChGs) method is expected to be applicable to these clouds.

The issue of whether the small scale structures in neutral hydrogen velocity channel maps are dominated by density or velocity structures has been debated recently, preprints by Clark et al. (2019)$^{\color{blue}{53}}$ and the response in Yuen et al. (2019)$^{\color{blue}{54}}$. However, irrespectively of the outcome of these debates, our conclusion that the velocity channel gradients trace magnetic fields well is not affected, especially in the regime of molecular clouds.
\subsection*{Sub-Block Averaging}
\label{sec:sub-block}

The use of sub-block averaging comes from the fact that the orientation of turbulent eddies with respect to the local magnetic field is a statistical concept. In real space the individual gradient vectors are not necessarily required to have any relation to the local magnetic field direction. Yuen \& Lazarian (2017)\cite{2017ApJ...837L..24Y} reported that the 
velocity gradient orientations in a sub-region--or sub-block--would form a Gaussian distribution in which the peak of the Gaussian fit reflects the {\it statistical most probable} magnetic field orientation in this sub--block. As the area of the sampled region increases,  the precision of the magnetic field traced through the use of Gaussian block fit becomes more and more accurate. Subsequently,  Lazarian et al. (2018)\cite{2018ApJ...865...46L} found that the width of the distribution is correlated with the statistical mean magnetization of the sub--region. The use of sub--blocks is a common feature of analyses like ours which use measured gradients in data to connect statistical theories of MHD turbulence to an understanding of gradient orientations in physical space. One should note that sub--block averaging is not just a smoothing method; it provides one with a new statistical measure of the data. Yuen \& Lazarian (2017)\cite{2017ApJ...837L..24Y} provide a detailed discussion of how white noise affects sub-block averaging {\it vs.}  common smoothing techniques.

\subsection*{Moving Window}
\label{subsec:MW}
Another technique developed to improve the performance of VGT is the  Moving Window (MW) method\cite{2018ApJ...853...96L}. The Moving Window method is an attempt to employ sub--block averaging in a continuous rather than a discrete manner. As magnetic fields are continuous, we move the block according to the orientation of the predicted magnetic field to smooth the outlying gradients. When there is an abnormal gradient vector compared to the neighboring vectors, we rotate the abnormal vector so that a smooth field line is formed. Mathematically, the rotation can be handled by performing smoothing on both the cosines and sines of the raw gradient angle, which is a convolution of an averaging kernel with the raw cosine and sine data. 

Previous studies show that there is a limit to how large one can make a Moving Window without the alignment being compromised\cite{2018ApJ...853...96L}. The size of the Moving Window chosen here is slightly smaller than the limitation, which not only improves the alignment between the orientations of gradients and dust polarization but also shows visually correct orientations of the gradients. We use 2 pixels as the MW width for Taurus, NGC1333, L1551,  Serpens, and Perseus A. For Smith Cloud, we choose 1 pixel as the MW width.

\section*{Observational Data }

\subsection*{Taurus} 
The Taurus Molecular Cloud\cite{2008ApJ...680..428G} region was measured in the $J = 1 - 0$ transition of $^{13}$CO using the 13.7m millimeter-wave telescope of the Five College Radio Astronomy Observatory (FCRAO). The data cover approximately
100 $deg^2$ of the sky (11.5$^\circ$ in R.A. by 8.5$^\circ$ in Dec.) corresponding to a region 28 pc $\times$ 21 pc at a distance of 140 pc. The high angular resolution of 47" allows one to examine in detail the relatively fine structures along with the large-scale distribution of the molecular material and the magnetic field. The RMS $1\sigma$ noise level is 0.18 K for $^{13}$CO in an individual pixel.

\subsection*{Serpens \& \textbf{NGC 1333}}
The Serpens cloud\cite{2013ApJS..209...39B}, which extends across a 50$'\times$60$'$ region corresponding to 5.3 pc $\times$ 7.6 pc at a distance of 415 pc, is a low-mass star-forming cloud in the Gould Belt, while NGC 1333\cite{2014ApJS..214....7B} is a 50$'$ $\times$ 60$'$ section of the Perseus Molecular Cloud (3.4 pc $\times$ 4.1 pc at a distance of 235 pc). The $^{13}$CO $J = 2 - 1$ emission data (220.4 $GHz$) on both regions were obtained by the Arizona Radio Observatory Heinrich Hertz Submillimeter Telescope. The angular resolution is 38" (0.04 pc) and velocity resolution is 0.3 km s$^{-1}$. The RMS $1\sigma$ noise level is 0.11 K for both data in an individual pixel.

\subsection*{Perseus A}
The Perseus molecular cloud\cite{2006AJ....131.2921R} is a nearby giant molecular cloud in the constellation of Perseus. The $^{13}$CO $J = 1 - 0$ emission data of Perseus A were taken from the COMPLETE Survey using the FCRAO telescope at an angular resolution of approximately 46". The RMS $1\sigma$ noise level is 0.15 K in an individual pixel.

\subsection*{L 1551}
L 1551\cite{2016ApJ...826..193L} is relatively isolated in the Taurus molecular cloud. Observations of the $J = 1 - 0$ transition of $^{13}$CO  were made using the Nobeyama Radio Observatory (NRO) 45 m telescope equipped with the 25-BEam Array Receiver System (BEARS) receiver. The data cover $\sim40'\times 40'$ with a resolution of $\sim$30", yielding maps with the highest spatial resolution. The RMS $1\sigma$ noise level is 0.94 K in an individual pixel.

\subsection*{Smith Cloud}
The Smith Cloud\cite{2013ApJ...777...55H} is one of the best high-velocity clouds for tracing the interaction between the Galactic halo and interstellar medium. It covers a  10.5$^\circ\times$  9$^\circ$ region at a  distance of 12.4 kpc. The H I data used here were obtained using the Rebert C. Byrd Green Bank Telescope. The spectra cover 700 km s$^{-1}$ around zero LSR velocity at a velocity resolution of  0.65 km s$^{-1}$ and an angular resolution is 9.1$'$, while the typical RMS $1\sigma$ noise level is 90 mK in a 0.65 $km s^{-1}$ channel.

\subsection*{Planck Mission}
Planck (http://www.esa.int/Planck) is a project of the
European Space Agency (ESA), with contributions from NASA (USA) and telescope reflectors provided by a collaboration between ESA and a scientific
consortium led and funded by Denmark. The Planck data we used here is Planck HFI Products for Public Data Release 3 2018\cite{aghanim2018planck}. The data is from the study of the polarized thermal emission from Galactic dust, using the High Frequency Instrument at 353 $GHz$ with angular resolution 5$'$. 
\subsection*{Data Availability}
The data that support the plots within this paper and
other findings of this study are available from the corresponding author and other co-authors upon reasonable request.

\newpage
\bibliography{reference}

\newpage


\section*{Acknowledgements}
AL acknowledges the support of the NSF grant AST 1715754, and 1816234, NASA grant NNX14AJ53G. PFG's research was carried out  at the Jet Propulsion Laboratory, which is operated for NASA by the California Institute of Technology. We acknowledge Mark Heyer for a number of valuable suggestions in improving our paper. We acknowledge the COordinated Molecular Probe Line Extinction Thermal Emission Survey of Star Forming Regions (COMPLETE) for providing a range of data for the Perseus and the Arizona Radio Observatory for providing the data of Serpens regions and NGC 1333. The Green Bank Observatory is a facility of the National Science Foundation operated under a cooperative agreement by Associated Universities, Inc. Based on observations obtained with Planck (http://www.esa.int/Planck), an ESA science mission with instruments and contributions directly funded by ESA Member States, NASA, and Canada. We thank the anonymous referees of the paper for many valuable suggestions.
\newpage
\section*{Author contributions statement}
All authors discussed the results, commented on the manuscript, and contributed to the writing of the manuscript. Yue Hu, Ka Ho Yuen, and A. Lazarian conceived the project,  Yue Hu, Ka Ho Yuen, and Ka Wai Ho performed calculations, while Yue Hu, Ka Ho Yuen, V. Lazarian, and A. Lazarian analyzed the results and wrote the original manuscript. Robert A. Benjamin provided suggestions on how this technique might be applied to the Smith Cloud. The data on the Taurus cloud were provided by Paul F. Goldsmith and those on the Smith cloud were provided by Alex S. Hill and Felix J. Lockman.

\newpage
\section*{Supplementary information}
\subsection*{Re-rotation Test}
Depending on the range of physical scales probed by the observations, self-gravity can be the main force affecting the dynamics of the gas in Giant Molecular Clouds (GMCs). We expect that the gradients to be parallel to the magnetic field with the presence of strong self-gravitational force \cite{2017arXiv170303026Y,2018ApJ...853...96L}. We test this by re-rotating the gradients by 90$^\circ$ again (thus called re-rotation) in the high-density region.

\begin{figure}[!h]
	\centering
	\includegraphics[width=0.90\linewidth,height=0.80\linewidth]{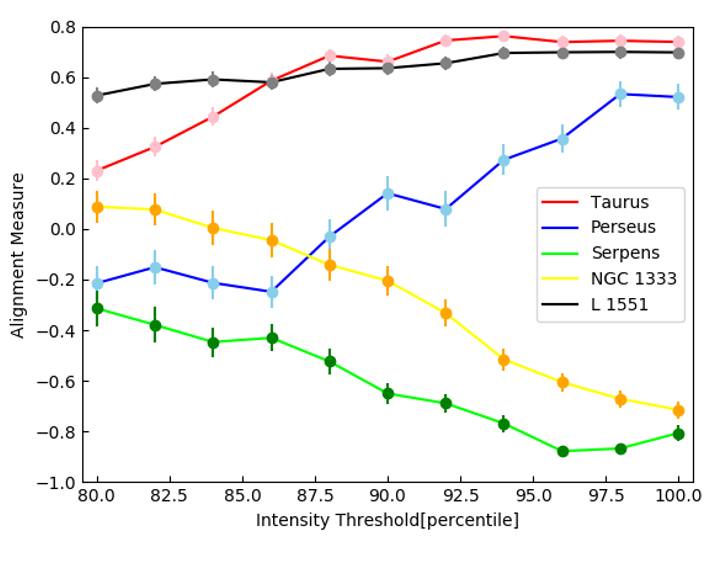}
	\caption{The variation of AM calculated over the velocity channel map with different re-rotation density threshold. Horizontal axis is the intensity threshold for re-rotation. When the intensity exceeds the threshold, we re-rotate the gradients in that region. The different coloured lines correspond to different clouds.}
	\label{fig:rerotation}
\end{figure}

\begin{figure}[!h]
	\centering
	\includegraphics[width=0.90\linewidth,height=0.80\linewidth]{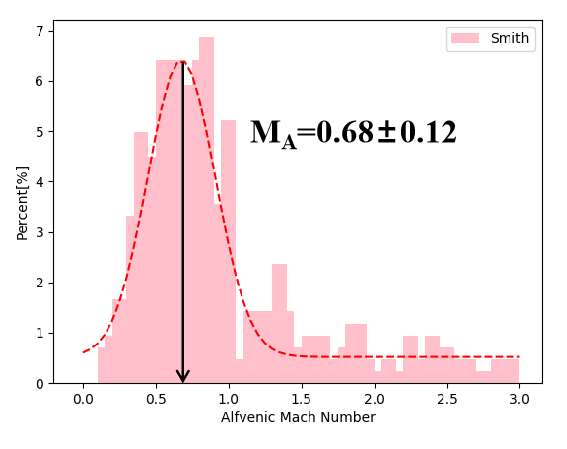}
	\caption{The histograms of the M$_A$ distribution obtained with VGT for Smith Cloud. The red dashed line represents the Gaussian fitting for the histogram. The vertical black line and M$_{A}$ indicate the expectation value of M$_A$ obtained from VGT for Smith Cloud.}
	\label{fig:smith}
\end{figure}

We define the concept of ``high density region'' by cutting off the density pixels below a certain threshold according to the percentile in the integrated velocity channel map. We use the same \textbf{Alignment Measure (AM)} to quantify the accuracy of the VGT when tracing the magnetic field as in the main text. We vary the threshold for the definition of the high-intensity region from the 80$^{th}$ percentile,  for which the AM is close to 0,  to the 100$^{th}$ percentile which means that there is no re-rotation. Supplementary Fig.~\ref{fig:rerotation} shows the AM variation with different re-rotated threshold. For Taurus\cite{2008ApJ...680..428G}, Serpens\cite{2006AJ....131.2921R}, L 1551\cite{2016ApJ...826..193L}, Perseus A\cite{2006AJ....131.2921R}, and NGC 1333\cite{2014ApJS..214....7B}, the maximum or minimum AM values are achieved at 100$^{th}$ percentile around which means no re-rotation. Thus, we conclude that the collapsing regions constitute only a small fraction in Taurus, L 1551, Perseus, while a large fraction in NGC 1333 and Serpens.

\subsection*{The strength of the magnetic field on Smith Cloud}
Supplementary Fig.~\ref{fig:smith} shows the histogram of M$_A$ derived from VGT in Smith Cloud\cite{2013ApJ...777...55H,1998MNRAS.299..611B,2008ApJ...672..298W}. The mean value of M$_A$ for Smith cloud is $0.68$, with a dispersion of $0.12$. We assume that the line-of-sight length is the transverse size of the cloud. This is actually an upper limit on the path length because the filling factor of the cloud may be less than unity. The lower limit on the number density $n$ can then be estimated $n > N / L > 5\times 10^{20} cm^{-2} / 1 kpc = 0.16 cm^{-3}$, where we chose the highest column density N through the cloud \cite{2008ApJ...679L..21L}. The volume density of the cloud is therefore $\rho = 1.4 \times n \times m_{HI} \sim 4*10^{-25} gcm^{-3}$ (the 1.4 accounts for the fact that the interstellar gas contains a fraction of Helium , $m_p$ is the mass of neutral Hydrogen). Assuming injection velocity $v_L = 10 km/s$, and using M$_A = 0.68$ as derived from the VGT, then $B = \sqrt{4 \pi \rho} v_L / M_A\geq 3 \mu G$.

In addition, Houde et al. (2009)\cite{2009ApJ...706.1504H} suggested that the strength of magnetic field estimated through the DCF method exits an amount of deviation $\sim\sqrt{\frac{B_0^2}{B_t^2}}$ due to the contribution of turbulent magnetic fields, assuming that the magnetic field $B$
is composed of a large-scale structured field $B_0^2$, and a turbulent component $B_t^2$, such that $B=B_0+B_t$. The same effect was addressed in terms of the eddy number along the line of sight by Cho \& Yoo (2016) \cite{2016ApJ...821...21C} using the fact that $\frac{\delta C}{\delta v_L}\sim \sqrt{\frac{B_0^2}{B_t^2}}$, where $\delta C$ is the dispersion of velocity centroid and $\delta v_L$ is the dispersion of velocity along line of sight. Replacing the $v_L = 10 km/s$ by $\delta C\sim 16km/s$ in Smith cloud, we obtain $B\geq 4.8 \mu G$.
\begin{figure*}[!h]
	\centering
	\includegraphics[width=0.99\linewidth,height=0.55\linewidth]{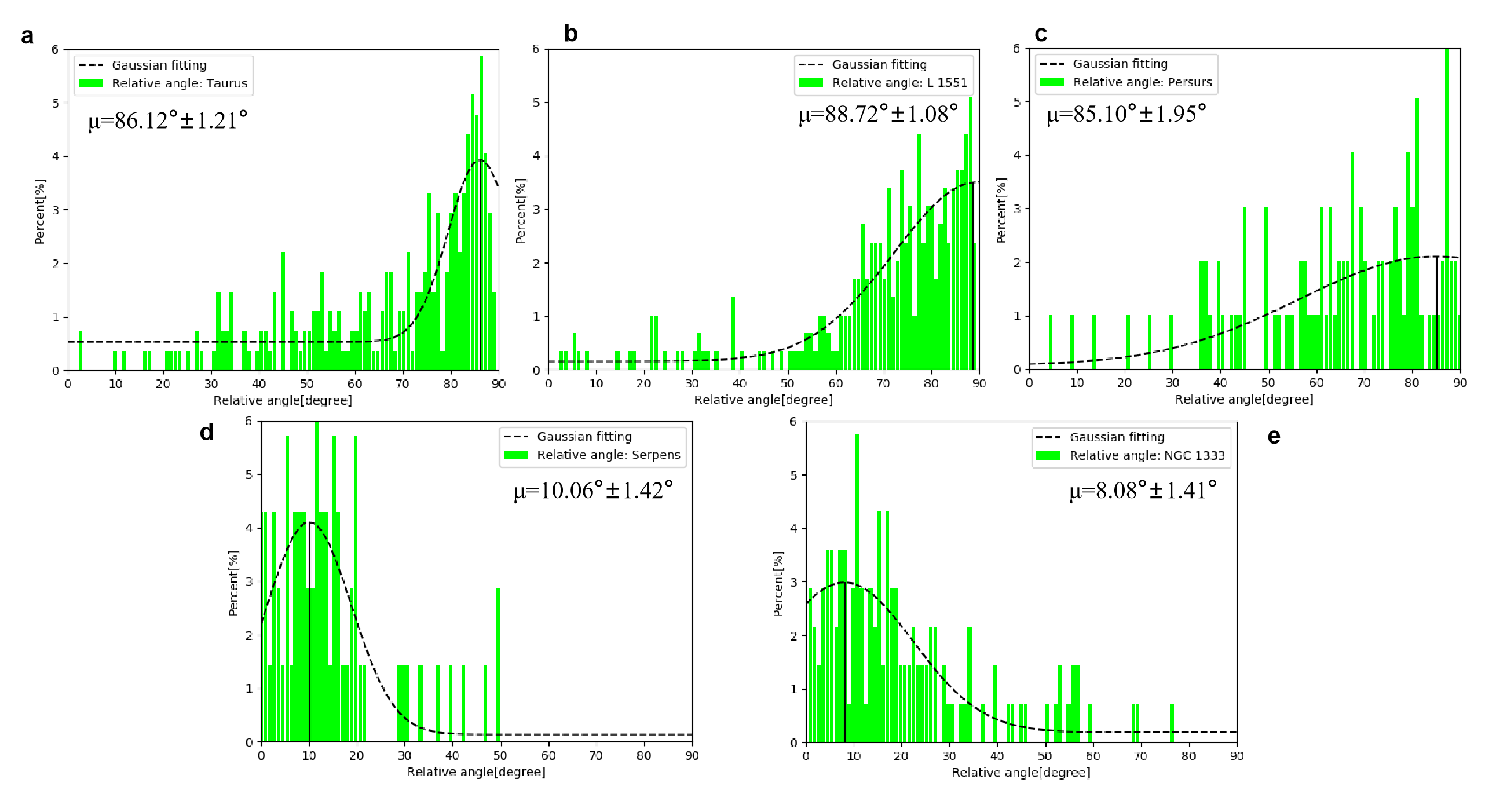}
	\caption{Distribution of alignments between re-rotated VChGs and the magnetic field inferred from Planck polarization. \textbf{a}, the histogram of relative orientation for Taurus. \textbf{b}, the histogram of relative orientation for L 1551. \textbf{c}, the histogram of relative orientation for Persus A. \textbf{d}, the histogram of relative orientation for Serpens. \textbf{e}, the histogram of relative orientation for NGC 1333. The distribution is drawn by using the gradient recipe after sub-block averaging \cite{2018ApJ...853...96L}. The dashed line is a Gaussian fitting to the distribution. $\mu$ is the expectation of the relative angle between the un-rotated gradients and the magnetic field derived from Planck polarization.}
	\label{fig:HRO}
\end{figure*}

\subsection*{Histogram of the relative orientation}
In Supplementary Fig.~\ref{fig:HRO}, we plot the Histogram of the Relative Orientation\cite{2013ApJ...774..128S} between non-rotated VChGs and the magnetic field inferred from Planck polarization. The distribution is drawn by using the gradients after sub-block averaging. For Taurus, Perseus A,  and L 1551, their peak values are very close to the theoretical value 90$^\circ$, while for Serpens, the deviation is 10$^\circ$. We do see a deviation of the theoretically predicted value (90$^\circ$) and that in observations between the gradients and dust polarimetry , especially for Serpens.  We list possible reasons for such deviation here.

Considering the deviation in the histogram, the noise in Stokes Q and U does not critically affect the shape of the histograms\cite{2016A&A...586A.138P}. However, the foreground and background will contribute to the deviation. The total Stokes parameters Q and U measured in each region can be interpreted as: Q=Q$_m$+Q$_b$, U=U$_m$+U$_b$, where Q$_m$ and U$_m$ correspond to the polarized emission from the molecular clouds, Q$_b$ and U$_b$ correspond to the polarized emission from the background. The contributions from the background polarized emission and the noise are estimated using the RMS of the Stokes parameters in the same reference region, Q$_{rms}$ and U$_{rms}$\cite{2016A&A...586A.138P}. In this paper, when we select the pixels to calculate the gradients, we are focusing on the structure of the molecular clouds in which Q$_m>>$Q$_b$ and U$_m>>$U$_b$ in order to minimize the contribution from the background. In addition, the sub-block averaging method, which is used to increase the reliability of important statistical measure in a region, also suppresses the contribution from the foreground and the background.

The histogram binning process can cause deviations\cite{2016A&A...586A.138P} in the histogram. For example, the histogram of Taurus shows the smallest uncertainty because of the large number of samples in each histogram bin, while L 1551 shows larger uncertainty. We found that the noise is the largest from the CO data on L 1551. As a result, we expect that there is a larger deviation between the gradients and its Planck measurements.  We also find systematic deviations between the polarimetry data from both Planck Collaboration VII 2015 PR2 353 $GHz$ polarization data\cite{2016A&A...594A...7P} and Planck Collaboration III 2018 PR3 353 $GHz$ polarization data\cite{aghanim2018planck}. Since the noise level for Planck 2018 is significantly smaller than that in 2015, we shall adopt the measurements from 2018 but with a caution that a deviation between them is possible. In Supplementary Fig.~\ref{fig:AM}, we plot the distribution of the alignment measure between the rotated VChGs and the magnetic field inferred from Planck polarization. This is helpful in spotting the regions in the maps where the alignment between the VGT and polarization show agreement or disagreement. The magnetic field vectors derived from both CO data and Planck data have already been smoothed into the same angular resolution after the sub-block averaging method.
\begin{figure*}
	\centering
	\includegraphics[width=0.99\linewidth,height=0.65\linewidth]{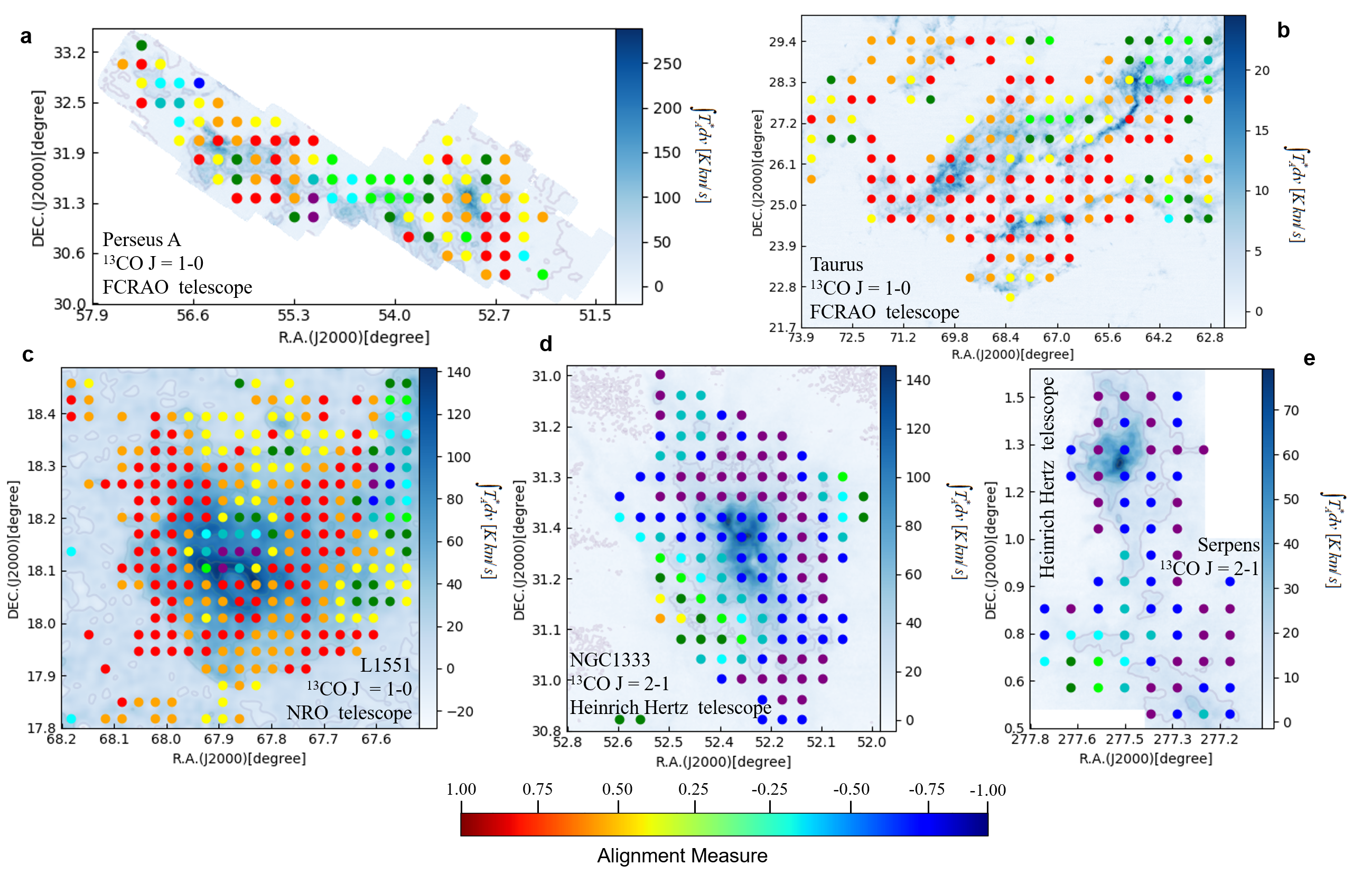}
	\caption{Distribution of the alignments measure between rotated VChGs and the magnetic field inferred from Planck polarization. T$_A^*$ is the antenna temperature. \textbf{a}, the distribution of the alignments measure for Perseus A. \textbf{b}, the distribution of the alignments measure for Taurus. \textbf{c}, the distribution of the alignments measure for L 1551. \textbf{d}, the distribution of the alignments measure for NGC 1333. \textbf{e}, the distribution of the alignments measure for Serpens.}
	\label{fig:AM}
\end{figure*}
\begin{figure*}
	\centering
	\includegraphics[width=0.99\linewidth,height=0.30\linewidth]{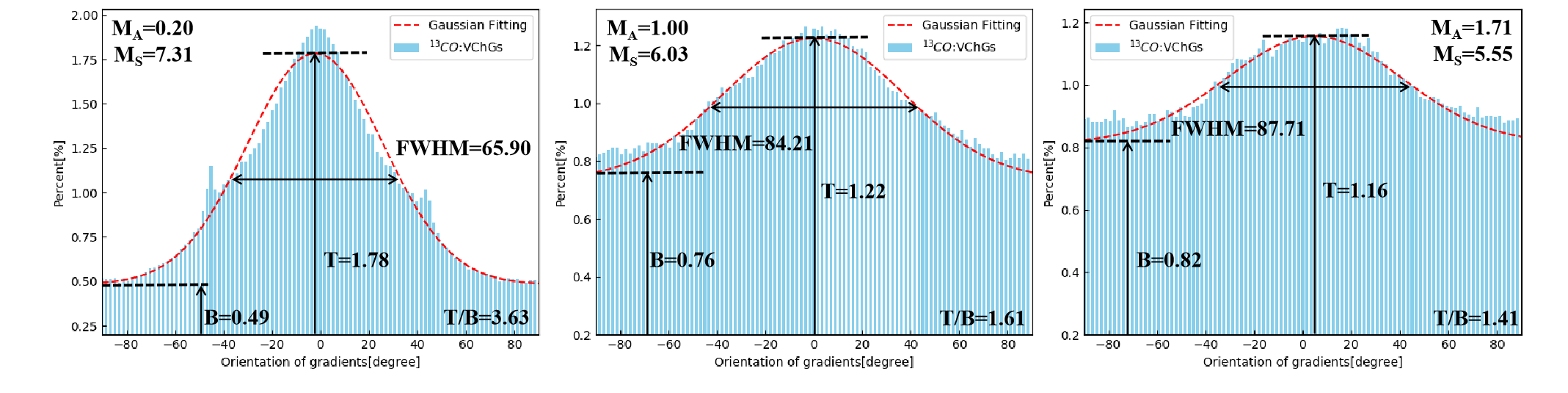}
	\caption{The normalized histograms of velocity channel gradients (VChGs) orientation from three numerical cubes with different Alfvenic Mach number M$_{A}$. \textbf{a,} the histograms obtained from synthetic observation with M$_A$=0.20, M$_S$=7.31. \textbf{b,} the histograms obtained from synthetic observation with M$_A$=0.40, M$_S$=6.03. \textbf{c,} the histograms obtained from synthetic observation with M$_A$=1.71, M$_S$=5.55. Three numerical simulation cubes used here have resolution 792$^3$. The fitted Gaussian is shown with the dashed red line. The T and B values (shown in each panel and noted with horizontal dashed black lines) the peak and base value of the Gaussian profile. FWHM = $2.355\sigma$ is the Full Width at Half Maximum, where $\sigma$ is the dispersion of the Gaussian profile.}
	\label{fig:dispersion}
\end{figure*}
\subsection*{ Obtaining $M_A$ from the distribution of gradient orientation}
In Fig.~\ref{fig:dispersion}, we provide a brief description of the gradient dispersion and how this can be used for deriving M$_A$. We use three super-sonic MHD simulations with different M$_A$ and apply radiative transfer code SPRAX \cite{2019ApJ...873...16H} to them (see Lazarian et al. (2018) \cite{2018ApJ...865...46L}for details about the numerical cubes). The Supplementary Fig.~\ref{fig:dispersion} shows the normalized histograms of velocity channel gradients (VChGs) orientation. We find that the width of Gaussian profile increases with respect to M$_A$, while the Top-Base (T/B) ratio decreases with the increase of M$_A$. Lazarian et al. (2018)\cite{2018ApJ...865...46L} showed there is a clear power law between gradients dispersion/Top-Base ratio and the M$_A$. Hence, by estimating the gradients' dispersion in each sub-block, we can obtain the M$_A$ in that region using the formula obtained in Lazarian et.al (2018)\cite{2018ApJ...865...46L}:
\begin{equation}
\begin{aligned}
	\frac{T}{B}\propto M_A^{-0.46\pm0.18}, M_A\le0.92\\
	\frac{T}{B}\propto M_A^{-0.25\pm0.02}, M_A>0.92\\	
\end{aligned}
\end{equation}

\end{document}